# Decision Making guided by Emotion

## A computational architecture

Dominique G. Béroule
Laboratoire d'Informatique pour la Mécanique
et les Sciences de l'Ingénieur,
LIMSI–CNRS UPR 3251, Orsay, France
dominique.beroule@limsi.fr

Pascale Gisquet-Verrier
Centre de Neuroscience Paris Sud, CNPS
CNRS UMR 8195, Université Paris-Sud
Orsay, France
pascale.gisquet@u-psud.fr

*Abstract*— **A computational architecture is presented, in which "swift and fuzzy" emotional channels guide a "slow and precise" decision-making channel. Reported neurobiological studies first provide hints on the representation of both emotional and cognitive dimensions across brain structures, mediated by the neuromodulation system. The related model is based on Guided Propagation Networks, the inner flows of which can be guided through modulation. A key-channel of this model grows from a few emotional cues, and is aimed at anticipating the consequences of on-going possible actions. Current experimental results of a computer simulation show the integrated contribution of several emotional influences, as well as issues of accidental all-out emotions.**

## I. INTRODUCTION

During its long-lasting evolution, the brain had to deal with a main physical constraint undergone by signal processing devices. Sensory systems are notably challenged by Uncertainty Principles in *Fourier Analysis* [1], according to which: *"the more a given analyzer locates a signal in the time domain, the less it can locate it in frequency/space and vice versa"*. In other words, setting up a sensory analyzer faces the following alternative: making it either "swift and somewhat ill-sighting" (like the peripheral vision), or "rather slow but sharp-eyed" (like the central vision). For surviving the primitive world predators, better being fast. This is why the ancient brain probably took up the first alternative, still operating today, on top of which in-depth but slower analyses have then developed as the environmental pressure decreased. The modern brain works with several parallel analyses of its environment, each tuned to a particular physical dimension, thus coping with the "uncertainty principle". The question of the interaction between these analyses then arises, as well as the possible generality of this perceptive architecture, e.g.: Could this sensory split in complementary channels extend to higher-order cognitive processes, as proposed in "Dual-system" models of decision-making [2]?

This question will be addressed in the first section, with references to neurobiological data: evidence coming from the literature suggests that several brain regions, as well as neuromodulation systems, are busy selecting proper behaviors.

Beside the main stream of Artificial Neural Networks (ANN), Guided Propagation Networks (GPN) rely on inner flows of activity which are controlled by neuromodulation-like processes. This main feature guided our choice towards their use in the present study, given the recognized link between emotion and neuromodulation. Following a reminder of the GPNs main characteristics, a specific architecture is gradually introduced in the second section. A software implementation of this model has been run on the same training data with different types of internal modulation and input cues. Related results are reported in the last section.

## II. CEREBRAL BASIS OF DECISION MAKING

### A. *Emotional marking of memory traces*

For living organisms, dangerous situations must be identified as quickly as possible for an appropriate response to be elicited. In plants, a water deficit triggers chemical signals inducing the inhibition of photosynthesis, among other metabolic changes. For organisms owning brains, inner warning signals can even be delivered long before the occurrence of a threatening event. This happens thanks to the memory function, which keeps track of experiences leading to meaningful events. In order to anticipate dangerous situations, the uncertainty principle suggests that associated cues should better be memorized in their fuzzy rather than high-definition format. These so-called emotional cues seem more efficient, due to their swift recognition. Such a signal processing constraint is consistent with the fast warning circuit aroused by the *amygdala*, a tiny brain structure of the ancient *limbic system* [3]. Fast decisions may involve emotions through physiological signals associated with their past outcomes, as proposed by the *Emotional Markers* hypothesis [4]. The functional part of emotion would include the enhancement of the behaviors memorized as a rewarding experience. Conversely, actions having previously led to unpleasant events would likely be avoided. Summing up, the emotional system can be viewed as a background "swift channel" fed by fuzzy cues which have acquired an emotional salience, meant to influence the decisions made by slower and more precise channels.

Within a topological memory device aimed at decision-making, can we expect decisions to only rely on markers fixed at a particular acquisition time? In this case, among several patterns of activation then aroused, the negatively marked items would always be avoided in favor of those associated with a positive *valence* (emotional value). However, as the same cue may be followed by situations of opposite valences, decision-making should account for several on-going factors. Moreover, thanks to their ability to memorize events developing over large time-spans, human-beings can project in a distant future. If every negatively marked situation were systematically avoided, nobody would bear disagreements possibly leading to delayed benefits. Who would dare writing a scientific paper if not motivated by the perspective of either being nobelised one day, or more likely enjoying a conference backstage in Australia? The emotional value of the uncompleted paper cue may then take opposite valences, depending on the enhanced time scale, which may vary as a function of the writer's mood at the moment the draft enters her/his peripheral vision. Again, the valence of a memory trace cannot be accurately determined in advance, but has to be established at the time of memory retrieval, instead.

*B. Emotional dimension across brain structures*

Decision-making is a complex process supported by the prefrontal cortex in which different roles are played by subregions, namely - orbitofrontal (*OFC*), anterior cingulate (*ACC*) and dorsolateral prefrontal (*DLPFC*). The *OFC* would notably code for the reinforcing value of the diverse stimuli encountered by the organism, and be involved in guiding action [5]. It would for instance be central for inhibiting responses that already led to undesirable consequences, and for enhancing an appropriate behavior despite the stronger activation provided by alternative ones [6]. The guiding part played by this cortical structure, possibly in association with the *ACC* [7], would favor actions that revealed rewarding in the past experience.

From a more general point of view, decision-making involves different brain regions working together to evaluate environmental stimuli and transform this information into appropriate actions. Accordingly, this process requires skills in cognition, emotion, memory and motor control, all carried out by structures in interaction with *reward circuits*. For instance, it has been shown that the emotional reactivation provided by retrieval cues facilitates the retention performance, and hence decision-making. These cues strongly activate the *amygdala*, a brain structure coding for emotion, as well as the *nucleus accumbens* which is a part of the reward system [8]. Recent literature in rodents, monkeys and humans is consistent with the idea that a *cortico-basal ganglia* circuit constitutes the heart of the reward processing. This circuit works with the retrieval-oriented *DLPFC*, and the emotion-oriented *amygdala* associated with regulating structures releasing neuromodulators such as *dopamine* (from the *ventral tegmental area*) and *serotonin* (from the *dorsal raphe*). Such *cortico-striato-thalamo-cortical loops* have a dual organization that permits both parallel and integrative processing. More precisely, as depicted in Figure 1, fibers from different prefrontal areas converge within subregions of the *striatum*: through the organisation of the *striato-nigrostriatal* projections, the ventral striatum can influence the dorsal striatum. The non-reciprocal cortico-thalamic projection conveys information from reward-related regions, through cognitive and motor controls. The initial discovery that the ventral striatum funnels emotional information towards motor outcomes has been investigated in non-human primates for building the *Ascending Spiral Model* [9][10]. Such a view is close to the one supported by brain imaging studies in human, which postulate that cortical regions are functionally linked through a "cascade" of interactions [11]. According to this model, three main loops (limbic, associative and sensori-motor) form a spiral-like pattern between the cortex and the striatum. Each step would comprise two parallel projections from a given cortical region, respectively towards its dorsal and ventral striatum partner. While the dorsal striatum would send an inhibitory feedback response to its cortical afferent, the ventral striatum would "*facilitate transfer of information to the next step in the spiral*", by modulating the adjacent cortical region. The latter would follow the same track through the striatum to influence its neighbour cortical region, and so forth until the final motor cortex (Fig. 1).

Such a system allows the anticipation of both possible rewards and punishments to favour a proper behaviour. The related facilitation or inhibition can be obtained by different means. These effects may involve specialized cortical regions: within the *OFC*, rewards tend to activate the medial part, while punishments seem to activate more lateral regions, instead. The dopamine, which is largely involved in the reward system, may exert opposite effects through a possible connectivity pattern proposed in the *Ascending Spiral Model* (Fig. 1). Finally, dopamine release may also induce opposite effects depending on the type of binding receptors. Although the main function exerted by dopamine appears to be mediated by inhibitory *D2* receptors, experiments also report the critical role of arousing *D1* receptors in reward-related learning [12].

*C. Impairment of emotional circuits*

Whereas most of our everyday choices do not cross our level of consciousness, patients with *prefrontal* damage lack this intuitive parallel evaluation of all possible options [13]. These patients are unable to perform optimal and rational decision-making. They miss the ability to anticipate the long-term consequences of their decisions, and preferably choose actions with the best short-term outcomes. Severe decision-making impairments are always associated with lesions of the prefrontal areas (*OFC*, *ACC* and *DLPFC*), an effect also observed in rat, probably due to a disruption of the basic modulating function [14].

A unifying view of two pathologies linked with emotion has recently been proposed, according to which exposure to extreme reinforcers of behaviour such as a trauma (very negative) or a drug of abuse (very positive) may both induce hypersensitivity to retrieval cues. This change of sensitivity would be responsible for drug craving in addiction, intense reviviscences in Post Traumatic Stress Disorder (*PTSD*), as well as relapse in these two pathologies [15].

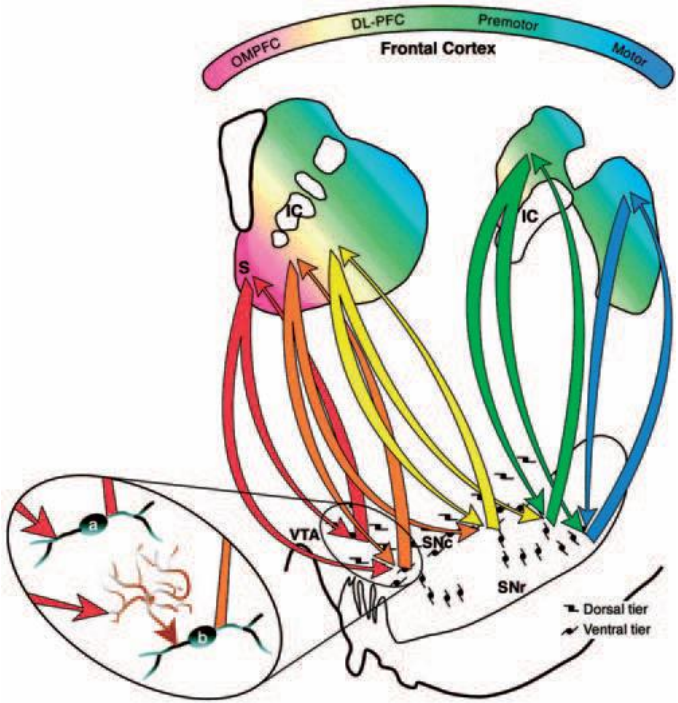

Figure 1. Ascending spirals of connectivity of the striatum to midbrain (downwards) and prefrontal cortex (upwards) (Reproduced from [9]). Magnified oval region to the left shows hypothetical cortico-striatal connections. The upper projection activates directly a dopaminergic cell (a) in the Dorsal tier, resulting in inhibition. The lower projection terminates indirectly on a dopaminergic cell (b) of the Ventral tier via an interneuron, resulting in facilitation. (VTA: Ventral tegmental area, OMPFC: orbitomedial perfrontal cortex, DLPFC: dorsolateral prefrontal cortex).

### III. NETWORK ACTIVITY GUIDED BY EMOTION

At the origin of ANNs, the *Perceptron* was already aimed at decision-making, namely for the categorization of visual patterns [16]. As this research stream enlarged with other formalisms, a great demand arose for aiding professionals in their decision tasks. Satisfying this demand has first been addressed with computer tools close to the initial ANN architectures [17], then incorporating functional principles motivated by psychological data linked with research on emotion [18]. More recent developments have followed advances in the physiology of emotional circuits of the brain, including elements of neuromodulation [19].

#### A. Sprouting of memory-pathways within modules

A Guided Propagation Network (GPN) is a modular architecture. Depending on its location within a matrix of channels and layers, each GPN module is focused on a particular representation of events (e.g.: typed words; Fig. 2). The same generic mechanisms are implemented in every module, all based on a main feature: an **inner flow of activity**. This spontaneous activity needs to be fed by a series of incoming stimuli for being able to cross the module and eventually reach its output where Feature-Detectors (FD) stand. Both rhythm and direction of the inner flow are given step by step while a given input pattern develops in time, leading to the full activation of the pattern-FD at the module output. The joint occurrence of the inner activity and input serial patterns implies that, at any moment, the front of the inner flow indicates the next stimuli to possibly occur, hence its predictive power.

In order to enable an unexpected stimulus and the inner flow to "hold future meetings", a new Elementary Processing Unit (EPU) can be brought into play in the course of processing (see Fig. 2). When a stimulus is found which did not participate in guiding the module inner flow, a EPU is taken from a pool, and assigned the role of *coincidence detection* between the stimulus and the current front. The memory location of a new EPU is thus given by the inner flow, provided the latter is focused enough. Little by little, chains of EPUs are created all along the network life by a non-supervised learning algorithm [20], forming pathways comparable to the branches of a tree. FDs standing at the end of the branches may guide the activation of more central modules. In other words, GPN-tree seeds are all sowed at once from the end of a given branch towards the ramifications of another module below. This is a waste of energy, but there is no time-loss, hence the theoretical real-time processing of GPNs.

#### B. Guided propagation through modulation

In a former GPN-model of reading [21], both precise perception of letters and scanner jumps were driven by the 'fuzzy and swift' perception of the text ahead of the letter currently fixed; the vocabulary subset sharing the same global shape could thus be enhanced. Despite its early response, this gross analysis was not to interfere with the on-going slow recognition process held in the neighboring channel. The solution taken on had been for a swift channel to decrease the decision thresholds of the EPUs participating in the representation of anticipated patterns. The upstream *facilitation* of EPU pathways has then been incorporated into the formalism, which incidentally opened the way to pattern production [22].

Any given GPN module has therefore the potential for either: 1-acquiring, 2-identifying or: 3-generating time-space patterns in several application fields [23].

A *Modulation device* is responsible for setting on-line the EPU parameters, which determine the module function among the aforementioned three ones. While a module is fed by a stimulus, its inner flow may thus be facilitated by upstream modulation signals (Fig. 3).

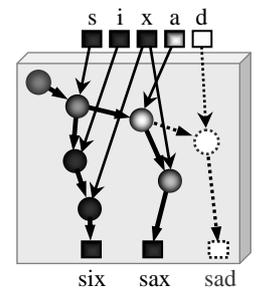

Figure 2. GPN module coding for words on the basis of one letter after the other (at the top). Two pathways (chains of EPUs: round cells, in dark) have already been built by the learning algorithm, when the following letters occurred for the first time, in succession:
- 's' at time $t_1$, 'i' at $t_1+\Delta t$, 'x' at $t_1+ 2\Delta t$ ;
- 's' at $t_2$, 'a' at $t_2+\Delta t$, 'x' at $t_2+ 2\Delta t$.

Now, 'a' has just occurred after 's', but the 'd' Feature-Detector (FD, square cell) has just fired instead of the expected 'x'. A new pathway is thus being created (dotted items), ending with a new module output: the labeled '**sad**' FD (at the bottom-right corner).
The inner flow originating from the top-left round cell can then be guided towards one of the current module FDs ('six', 'sax', 'sad'). The part of these word-FDs compares to the letter-FDs one, since the former may guide the inner flow of a more central module (not plotted here).

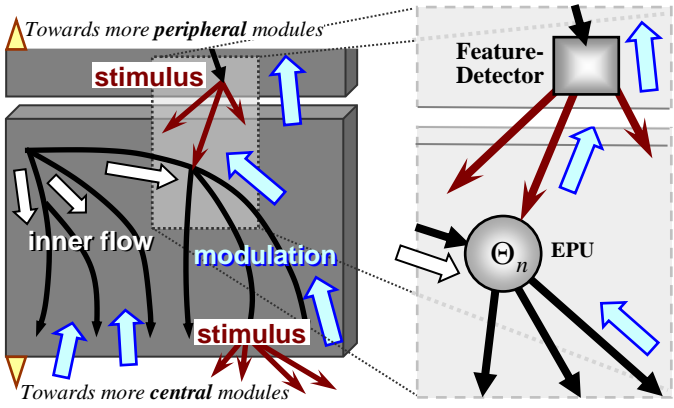

Figure 3. Content of a GPN module. In the left image, a trained module contains 'inverted tree'-like pathways, fed by an inner flow. This awaiting stream is guided by 1/ series of stimuli from more peripheral modules and 2/ modulation from other modules (upward arrows). In the magnified window to the right, an Elementary Processing Unit (EPU) owns a threshold $\Theta_n$ for detecting the possible juxtaposition of its two inputs: stimulus and inner flow.

The EPU standing at the "meeting spot" of the module incoming signals, is activated above its decision threshold $\Theta_n$, expressed as a fraction of the EPU maximum input (where $A$ is the maximum signal intensity, $E_n$ the EPU *Excitability* and $R_n$ the weight of the "Context input" fed by the inner flow). Above its threshold, the EPU output is proportional to its summed input, up to the $A$ saturation value.

$$\Theta_n = \frac{(1+R_n) \times A}{E_n} \quad (1)$$

Although a module inner flow has a pace usually made by current incoming stimuli, it can also be induced to run ahead the present time. This is a precious asset in the framework of this study, for a module to be able to anticipate salient cues conveying an emotional valence. The inner flow is then reinforced by shifting the parameters introduced in formula (1) from the "restricted" to the "extended" mode (see Fig. 7), and this for every EPU of its become "far-sighted" module.

The logic behind the resulting multiple-channels architecture can be expressed in this way: the "aware" channel (*Decision Maker*, labeled *III* in Fig. 5) keeps pace with present stimuli, including *proprioceptive* ones from actions being performed (channel *IV*). In order to quickly select the future right action among possible ones, channels *III* and *IV* need to be guided by modulation signals (Fig. 4) issued from another channel (*Cues Binder*, *II*) in which the pathways of future events are activated in advance. The respective emotional values of these anticipated possible events need to be given by another channel (*Emotion Detector*, *I*).

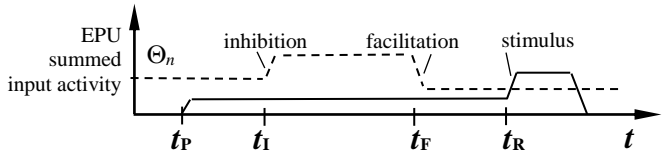

Figure 4. Instance of EPU activity diagram. This EPU is primed by the inner flow at time $t_P$. Its decision threshold (dotted line) then undergoes two modulation signals: an increase (inhibition) at time $t_I$, a decrease (facilitation) at time $t_F$. Thanks to this strong facilitation, the EPU treshold is more likely to be crossed at time $t_R$, when a stimulus meets the inner flow.

## C. Conditioning through coincidental module outputs

In a simulated "ascending spiral" architecture, channels *I* and *II* only respond to fuzzy cues, whereas *channel III* is activated by well-defined stimuli. Quite a few emotional cues initially feed the *channel I*. The recruitment of new emotional stimuli permits decision-making to be guided in a more subtle way. Through the sprouting of a new pathway in *channel II*, any neutral fuzzy stimulus can be paired with an emotional cue occurring shortly after it. In *channel III*, another pathway is simultaneously built, which receives the precise version of both stimuli (Fig. 5).

When the behavior begins with a neutral stimulus (e.g.: the bell in Pavlov's dog trial), and ends with an emotional cue (i.e.: the sight of meat, or Unconditioned Stimulus *US*), three parallel pathways are activated in their respective GPN channels, which results in cross-connections between them. This acquired connectivity pattern is aimed at conveying modulating signals from one module to its right-hand side neighbor, forming the ascending spiral. The *channel I* outflow selectively facilitates *channel II* pathways: each repetition of a sequence ending by an emotional cue increases the inner flow along a pathway of *channel II*, making it draw ahead the current input. In this way, a neutral stimulus becomes "conditioned" (*CS*), i.e.: able to fully activate a *channel II* pathway as if the *US* had actually occurred (e.g.: in the Pavlov trial, the dog salivates as soon as the bell rings). This schema applies to aversive conditioning as well, through which moving towards cues associated with stressful events is prevented. As far as the '*II=>III*' cross-connection is concerned, the emotional value aroused by a *channel II* output induces the modulation of propagation within *channel III* (Fig. 5).

In order to implement 2nd order conditioning (e.g.: a red light preceding the conditioned bell sound induces salivation),

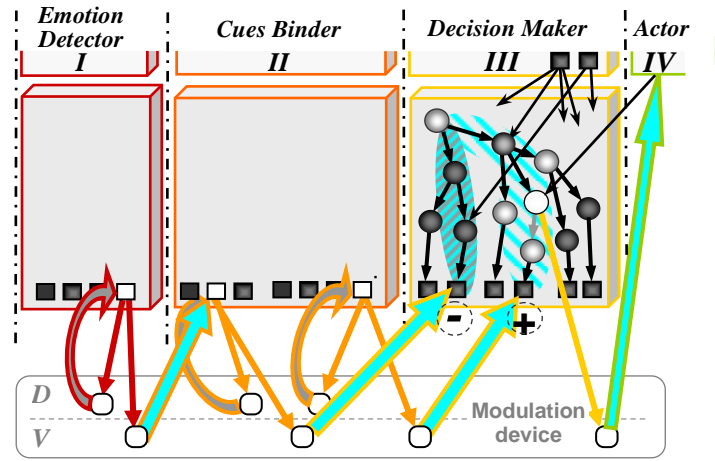

Figure 5. Representation of the potential interactions between GPN modules via a Modulation device (at the bottom). Both connectivity and color conventions partly come from Fig. 1. Downward straight arrows issued from the modules outputs stand for the activation of *D* and *V* tiers, the respective functions of which compare to the structures shown in Fig. 1: whereas *D* is in charge of inhibitory feedbacks (upward curved arrows), *V* sends modulating signals towards the next neighbor module to the right-hand side (upward left-to-right arrow). Modulation is either 'facilitating' (**+**) or 'inhibiting' (**-**). The modulating signals propagate upstream through *channel III (striped areas)*, thus inducing one of the currently possible actions via the modulating device.

a stimulus which repeatedly triggers one of the *US* responses will join in the *channel I* input. This mechanism results in a "funnel effect", which tunes a subset of cues with the reward. Conversely, in aversive conditioning, "an umbrella is opened" under all the cues associated with a negative emotion, thus trying to avoid its revival. This capacity requires the parameters to be properly settled, so that a *channel II* pathway gets fully activated under the only influence of its initial stimulus.

In the GPN theory, the increment of the contextual weight ($R_n$ parameter) along a pathway accounts for the integration of the previous stimuli, that is: $\forall n, R_{n+1} = 1 + R_n$ (2)

The output $C_n$ of the $n^{th}$ EPU along a pathway (Fig. 6) is expressed above the EPU threshold as a linear function of the summed Stimuli $S_n$ and the weighted output $C_{n-1}$ of the previous EPU.

$$C_n = \frac{S_n + R_n \times C_{n-1}}{1 + R_n}$$
$$= \frac{S_n + R_n \times (S_{n-1} + R_{n-1} \times C_{n-2})}{(1 + R_n)(1 + R_{n-1})}$$

In which: $S_n = S_{n-1} = 0$ and $C_{n-2} = A$

$$C_n = \frac{R_n \times R_{n-1} \times A}{(1 + R_n)(1 + R_{n-1})}$$

For the internal flow to freely cross two EPUs without stimuli, the upper boundary of Excitability $E_{n+1}$ is such that the weighted output stays below the next EPU threshold (see (1)):

$$R_{n+1} \times C_n \leq \frac{(1 + R_{n+1}) \times A}{E_{n+1}}$$

$$\frac{R_{n+1} \times R_n \times R_{n-1} \times A}{(1 + R_n)(1 + R_{n-1})} \leq \frac{(1 + R_{n+1}) \times A}{E_{n+1}}$$

Taking (2) into account, one gets: $E_{n+1} \leq \frac{R_{n+1} + 1}{R_{n+1} - 2}$ (3)

For the inner flow to cross a single EPU, a similar rule can be obtained: $E_{n+1} \geq \frac{R_{n+1} + 1}{R_{n+1} - 1}$ (4)

Returning to the assumption underlying a GPN model of Parkinson's disease symptoms [22], the $R_n$ parameter shown in Fig. 7 would compare to *serotonin*, whereas $E_n$ would play a part akin to *dopamine* in the neuromodulation system.

### D. Merged time-scales for Decision-making

Drugs of abuse are associated with immediate rewards, and cause serious trouble in the longer term. In that case, anticipating social and personal degradations may not be strong enough to counterbalance drug seeking.

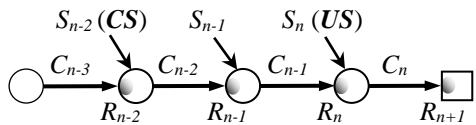

Figure 6. Instance of a *channel II* GPN pathway, showing its associated symbols ($S_n$: Stimuli, $C_n$: Contextual input, $R_n$: contextual weight)

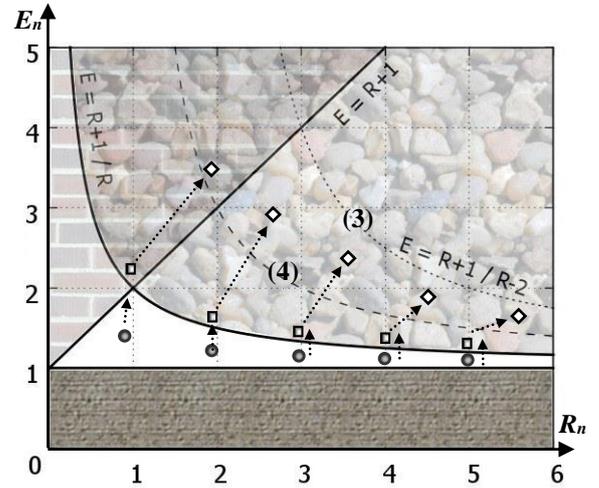

Figure 7 (adapted from [23]). Couple of EPU control parameters. The Excitability $E_n$ (vertical axis) and the inner flow input weight $R_n$ (horizontal axis) determine the behavior of a given EPU. In the "restricted propagation" area (in white), a newborn EPU requires both its Stimulus and inner flow (Context) inputs to fire. The distribution of round dots follows the required arithmetic progression of $R_n$ along a pathway of EPUs. In the (pebbly) "extended" propagation mode, $E_n$ and $R_n$ (square dots) allow older EPUs to fire thanks to the only inner flow. After having crossed the border between the "restricted" and the "extended" areas, several chained EPUs may get activated at once as the (4) and (3) conditions (see text) are satisfied (diamond shapes).

Conversely, an immediate effort may later on be rewarded. This dual time-scale influence can be modeled by taking advantage of the hierarchical nature of GPNs. Two layers with independent modulating circuits can be brought into play; one (central) layer integrating the stimuli of a more peripheral one. Thus, both modules of *channel II* may guide decision-making, sometimes with opposite modulation values (see Fig. 8).

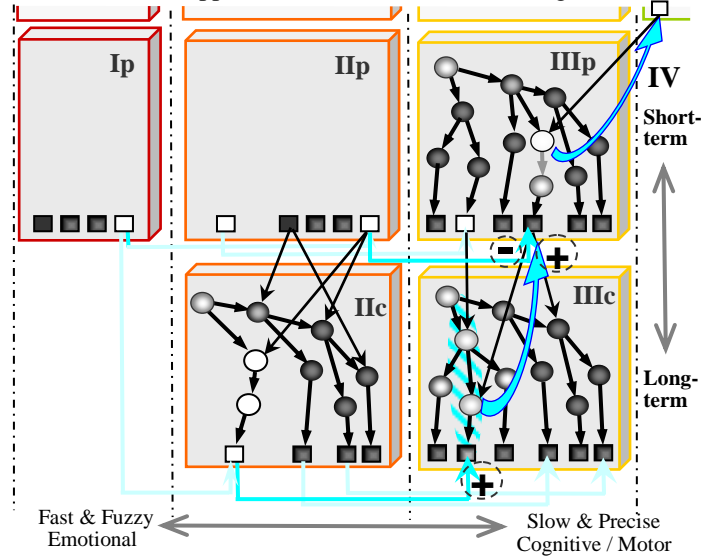

Figure 8. Full architecture of the model. The label of each module gives its location; e.g. **IIIp** corresponds to the 3$^{rd}$ channel and **p**eripheral layer, and **IIc** indicates the **c**entral module of *channel II*. The neuromodulation connectivity pattern plotted in Fig. 5 is reduced here to straight (blue) arrows at the modules bottom. The outflow of **IIc** is currently facilitating a pathway of **IIIc** (striped area), the EPU thresholds decrease of which is propagated upstream. As the **IIIc** inner flow follows this path, facilitation signals are sent towards a **IIIp** pathway (upward curved arrows). The latter is here inhibited at the same time by **IIp**, but not enough for avoiding the action selected by the facilitating flow issued from the central **IIc**.

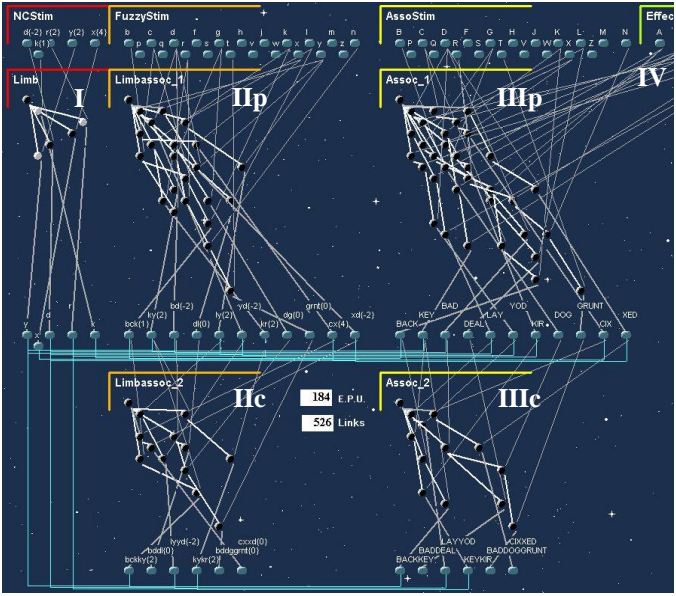

Figure 9. Screen shot during the training session. So far, the network has grown from the scanning of 6 combinations of 11 words, as shown by the current numbers of Feature-Detectors at the bottom of the modules.

## IV. EXPERIMENTS

Experimental issues are investigated on the same instance of GPN, obtained through a preliminary training session. The network is first set according to the global architecture shown in both Fig. 8 and Fig. 9, where five modules initially own a single EPU at the source of their inner flow.

The network interface depends on patterns to be processed. Here, a compound behavior is represented by a sequence of short words, in which stimuli (consonants) alternate with actions (vowels). The following Letter-Detectors are involved:
- A few unconditional stimuli or actions feed the *channel I* (i.e.: symbols 'd' of valence {-2}, 'r' {+2}, 'y' {+2});
- Twenty – initially neutral – stimuli (consonants) in their fuzzy format (i.e.: lower-case) at the top of *channel II*;
- Twenty stimuli in their high-definition format (i.e.: upper-case consonants) form the input of module IIIp;
- Six action effectors (i.e.: upper-case vowels) are ready to deliver *proprioceptive stimuli* after having generated the action triggered by module IIIp.

### A. Script of the experiments

An "elementary behavior" thus takes the shape of a short word. The software is reading a file, each line of which contains 2 or 3 such words (e.g.: 'BAD DOG GRUNT', 'BUZ ZARD'). While a file line is scanned, the concerned Letter-Detectors are activated one after the other at the network input.

Starting from nearly scratch, the software goes through the training of 100 combinations of 170 words which eventually brings 1200 EPUs and 4700 links in the current simulation. In the data, each first consonant is possibly followed by any of the vowels; in other words, after an initial cue, all available actions may be possible, among which the system will have to choose. This training period compares to the exploration of an environment comprising a few preset emotional cues; a series of actions in this environment may thus either be neutral or be concluded by a salient cue (i.e.: with valence {+/-2}).

When a new couple of words is read from the data file during training, memory pathways are sprouting in parallel inside every concerned module. Two elementary behaviors are memorized in **IIIp**, while **IIIc** grows a pathway combining them. The *channel II* only integrates a fuzzy version of the stimuli contained in these behaviors. For instance, the 'LOCK KEY' input results in the following pathways:

- in module **Ip** : one-EPU pathway labeled 'y'
- in **IIp** : 'lck' and 'ky'   / - in **IIc** : 'lckky'
- in **IIIp** : 'LOCK' and 'KEY' / - in **IIIc** : 'LOCKKEY'

An input ending with an emotional cue (i.e.: 'd', 'r', 'y') leads to new cross-modulation connections between modules outputs. As described in the previous section, these modulation links permit conditioning to operate in *channel II* with the help of *channel I*, as well decision-making in *channel III*, guided by channel II (Fig. 8). In this experiment, cues becoming "emotional" through 2$^{nd}$ order conditioning are assigned half the valence of their conditioning stimulus, namely either {-1} or {+1}. At the end of training, depending on the number of repetitions required for a cue to be conditioned (parameter to be fixed beforehand), several new FDs have been created in the *channel I* input. The software is run on the training data (which takes 4 minutes on an *Intel Pentium4* running at 1.5 *GHz* with 512 *Mo* of *RAM*). Given that new emotional cues appear in the course of training, some of them may have been created after the presentation of an input they could influence, hence a second presentation of the trial data.

During the following "test session", the system is given a cue (i.e.: a consonant), and is then free to decide what action to perform among the options previously met in the same context during training. The challenge for the **IIIp** inner flow is to be guided towards the most rewarding emotional target (Fig. 10).

It is worth noticing that less than 3% of the decision maker EPUs are firing simultaneously in the 'aware' *channel III* during this process, while around 6 times more EPUs are involved in the 'subconcious' *channel II*.

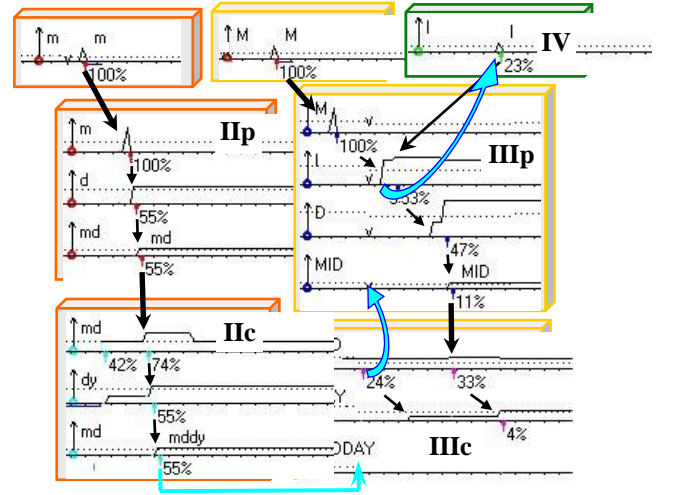

Figure 10. Activity histograms of pathways involved in guiding a decision through central modules (**IIc** and **IIIc**). An initial cue ('M') feeds channels *II* and *III*, channel *II* undergoes a swift cascade of activations across its two modules. Meanwhile, the EPUs linked with next possible actions are primed in **IIIp**. A facilitation wave (upward curved arrows) then selectively decreases *channel III* thresholds, making an effector ('I') fire in *channel IV*.

## B. An evaluation of each GPN module contribution

A *channel II* module can be withdrawn from the network either by switching off its inner flow generator, namely its root EPU, or by voiding its modulating output. Submitted to the same series of input stimuli, the output of the full architecture can thus be compared with the outcome of the system deprived of some modules. Among the alternative above, the weighting of modulation has been chosen for it offers the finest tuning of each module contribution to decision-making. $W_p$ is the weight of modulation between **IIp** and **IIIp**; $W_c$ holds the same part between **IIc** and **IIIc**. The results plotted in Fig. 11 are obtained with six different values of ($W_p$, $W_c$).

Tests are thus performed with these six types of modulation on 20 input cues (consonants). For each trial, this is the activity of the Effectors (*channel IV*) induced by a given initial cue which is considered significant. For example, 'B' may be followed by every vowel, including either 'A' or 'O' ('BAD DAY', 'BOY YARD'). The choice between the possible actions may be guided by both the peripheral modulation (**IIp->IIIp**) of the "word layer" and the central one (**IIc->IIIc**) of the "word-sequence layer". Given the opposite emotional valences of 'D'{-2} and 'Y'{+2}, the peripheral modulation favors the choice of 'O' because of the positive end of 'BOY' whereas 'A' is inhibited. On the contrary, the central modulation exerted via the last words cues ('DAY' and 'YARD') orientates decision-making towards the 'A' choice.

When both $W_p$ and $W_c$ are null, the consequent blocking of modulation drastically depresses the *channel III*. In absence of emotional information, the EPUs thresholds stay at their resting level, with only tiny differences reflecting the frequency information, namely the number of behaviors repetitions. This is why a general increase of the module EPUs Excitability can partly make up for the loss of action triggering, provided that one of the actions is more frequent than the others in the tested situation.

In current results, it appears that the anticipation of a close emotional cue has a modulating effect at least 3 times stronger than the prediction of a future one. This relative deficit of future cues can be attributed to the longer path taken by the central modulation (Fig. 10). This can also be interpreted in terms of "missing inhibition" from the deeper module **IIIc** towards module **IIIp**, given that in GPNs, only facilitating signals propagate upward within a given channel.

As far as *channel I* is concerned, switching it off after a training session does not impair the system global behavior, but prevents later emotional conditioning.

## C. A simulation of hypersensitivity to retrieval cues

During the training session, the valence of the initial emotional cues is set to +/-2, while conditioned stimuli have gained a valence of +/-1. The system can then be submitted to stimuli of extreme valence, which implies the one-shot conditioning of the cues that precede them in the input. Furthermore, the modulation power of these all-out stimuli is so high that the pathways they touch are either fully inhibited or fully facilitated in comparison with conventional emotional modulations (Fig. 11).

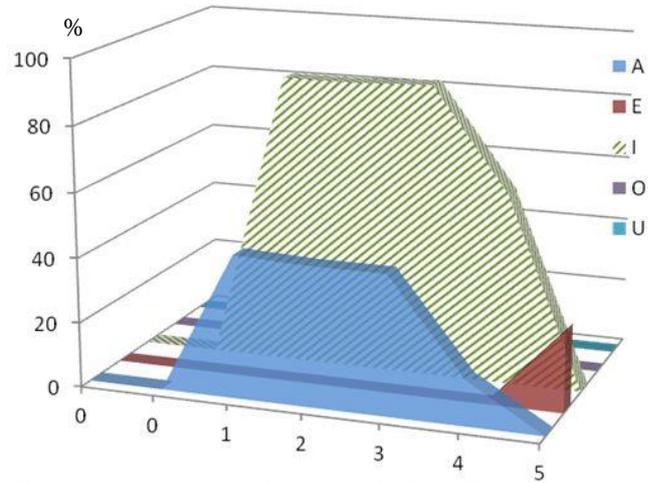

Figure 11. Prototypical activity rate of GPN effectors as the result of emotional training. During training, every initial input stimulus was followed by any of the possible actions ('A', 'E', 'I', 'O', 'U'), followed itself by either immediate or future stimuli owning emotional valences. During the "decision making" test, the system had then to choose which action to perform in response to a given input stimulus. This figure shows the effectors activity rate induced by the same input stimulus ('C'), as a function of six different combinations of peripheral and central modulation weights ($W_p$, $W_c$) taking the following values along the front axis: **0**:(0, 0); **1**:(1, 0); **2**:(1, 0.5); **3**:(1, 1); **4**:(0.5, 1); **5**:(0, 1). As with any input stimulus of the corpus, no decision is made without modulation (null response of all the action effectors at the origin of the front horizontal axis). Here, the central modulation is not amplified by a factor of 3 (see IV*B*). Consequently, this is only when the peripheral influence is switched off (tag '5'), that the expectation of a future reward becomes significantly efficient (activation of the 'E' effector at 25% of the maximum intensity). The greater response (90%) is obtained in the current context by the 'I' action, associated in the training data with the highest immediate drug-like reward. When the data does not include this highly rewarding stimulus, the figure obtained is close to this one ('A': 43%; 'E': 25%), but free of the 'I' response (striped area). When no "drug" is experienced during the training session, an effector barely produces a strong response, but the model simulation remains selective, even without lateral inhibition between its action-effectors.

Let us assume that the 'X' character be chosen as a highly rewarding cue associated with drug consumption. After a single occurrence of the 'SIX' input, 'S' enters *channel I* with a positive valence. Then, any series of actions leading to 'S' (e.g.: I and A in 'BIAS') will be favored, as well as other stimuli leading to the highly facilitated 'SIX'. This way, 'S', as other possible hypersensitized cues, participates in recruiting several other stimuli, possibly in one-shot. This looks as if conditioning could in this case enlarge the "receptive field" of an all-out stimulus, making the system more likely to meet in its environment a cue driving it towards the abuse of drug.

Assuming limited modulation resources, the involved strong – but local – facilitation of some *channel II* pathways may affect the Modulation device global balance. This system impairment will be studied in connection with the brain neuromodulators imbalance caused by drug addiction [24]. The treatment of compulsive drug-seeking and the avoidance behavior generated by a traumatism both evolve on a long-term scale, respectively leading to a possible relapse into addiction or revival of stressful situations. The assumption that similar mechanisms would cause the appearance of these different disorders will also be investigated with a new release of the computational model, in light of neurobiological data [15].

## V. Conclusion

Developed from a specific ANN, the model introduced here accounts for neurobiological evidence of the major part played by neuromodulation and emotional circuits in decision-making. The key component of the proposed architecture stands at the interface between the perception of emotional cues and modules oriented towards action. Referred to as "*channel II*", this component is able to chain fuzzy environmental cues, especially those which end by an emotional stimulus; it can then run ahead the present time and modulate the "decision maker" *channel III* so that its two internal flows are aspirated towards the most rewarding output, while triggering actions distributed along the preferred (path)way.

This architecture notably builds on the ascending spiral hypothesis linking the midbrain to the prefrontal cortex. A software implementation already confirmed the functionality of the involved mechanisms and allowed to start investigating their issues in the case of extreme behavioral reinforcers. As a representation of biological data, this model raises several expectations that could be considered later on, including: the hierarchical organization of emotional circuits, the greater (anticipatory) activation in subconscious emotional modules than in aware decision-making regions, the enlargement of emotional regions during a life-span, involving the creation of new cue-detectors, and the assumption that emotional consequences of a given situation be subconsciously replayed at every new instance of a similar situation. As a matter of fact, the reconstructive nature of memory is an important feature of retrieval processes, as often underlined since Bartlett [25].

Over a life span, the memory landscape is gradually shaped in our mind, including easy routes that are frequently visited. Meanwhile, in the background, a swift system keeps vigil at specific cues of the environment, those which acquired salience in the past, because associated with an emotion. Whenever such retrieval cues are detected, the current landscape would be temporarily reshaped with a combination of hard-to-climb hills and promising valleys. If cues previously associated with an extreme emotion occur, either a high peak or a ravine adds a sharp item to the local relief: the memory of a traumatic episode grows an unreachable plateau, while a drug of abuse digs an inescapable deep hole. Decision-making is therefore restricted, together with the freedom to explore the full landscape. Fortunately, if the theory presented in this paper is right, these strong landscape modifications may not be permanent, which would allow their modulation to be locally controlled, and the causative conditioned cues to become extinct: an opportunity to be grasped in future work.

## Acknowledgement

The study reported in this paper has been made possible thanks to Vincent Bloch who 30 years ago got interested in the first release of GPNs, Jean-Marc Edeline who more recently initiated the connection between the authors, and Vincent Douchamps who put the Ascending Spiral Model to the front. This work has been supported by a "NeuroInformatique CNRS" grant.